\documentclass[table]{aa_old} %

\usepackage[colorlinks=true,     linkcolor=blue, citecolor=blue, filecolor=blue, urlcolor=blue]{hyperref}
\usepackage{xcolor}
\usepackage{graphicx}
\usepackage{txfonts}

\usepackage{tabularx}
\usepackage{ulem}
\usepackage{gensymb,ulem}
\newcommand{\B}{B}
\newcommand{\Tp}{T_{\textrm{p}}}

\newcommand{\np}{n_{\textrm{p}}}

\newcommand{\vp}{v_{\textrm{p}}}

\newcommand{\ac}{a_{\textrm{col,p-p}}}
\newcommand{\nO}{n_{O^{7+}}/n_{O^{6+}}}
\newcommand{\lnO}{\log{n_{O^{7+}}/n_{O^{6+}}}}

\newcommand{\Osix}{O^{6+}}
\newcommand{\Oseven}{O^{7+}}

\begin{document}
\title{Scope and limitations of ad hoc neural network reconstructions of solar wind parameters}
\titlerunning{}
\author{Maximilian Hecht \inst{1}, Verena Heidrich-Meisner \inst{1}, Lars Berger \inst{1}, and Robert F. Wimmer-Schweingruber \inst{1}}

\authorrunning{M.H., V.H.-M., L.B., R.W.-S.}

\institute{Christian Albrechts University at Kiel, Germany,
  \email{heidrich@physik.uni-kiel.de}
}

\date{}

\abstract{
Solar wind properties are determined by the conditions of their solar source region and transport history. Solar wind parameters, such as proton speed, proton density, proton temperature, magnetic field strength, and the charge state composition of oxygen, are used as proxies to investigate the solar source region of the solar wind. The solar source
region of the solar wind is  relevant to both the interaction of this latter with the Earth's magnetosphere and to our understanding of the underlying plasma processes, but the effect of the transport history of the wind is also important. The transport and conditions in the solar source region affect several solar wind parameters simultaneously. Therefore, the typically considered solar wind properties (e.g. proton density and oxygen charge-state composition) carry redundant information. Here, we are interested in exploring this redundancy. 
}
{
The observed redundancy could be caused by a set of hidden variables that determine the solar wind properties. We test this assumption by determining how well a (arbitrary, non-linear) function of four of the selected solar wind parameters can model the fifth solar wind parameter. If such a function provided a perfect model, then this solar wind parameter would be uniquely determined from hidden variables of the other four parameters and would therefore be redundant. If no reconstruction were possible, this parameter would be likely to contain information unique to the parameters evaluated here. In addition, isolating redundant or unique information contained in these properties guides requirements for in situ measurements and development of computer models. Sufficiently accurate measurements are necessary to understand the solar wind and its origin, to meaningfully classify solar wind types, and to predict space weather effects. 
}
{
We employed a neural network as a function approximator to model unknown, arbitrary, non-linear relations between the considered solar wind parameters. 
This approach is not designed to reconstruct the temporal structure of the observations. Instead a time-stable model is assumed and  each point of measurement is treated separately. This approach is applied to solar wind data from the Advanced Composition Explorer (ACE).
The neural network reconstructions are evaluated in comparison to observations, and the resulting reconstruction accuracies for each reconstructed solar wind parameter are compared while differentiating between different solar wind conditions (i.e. different solar wind types) and between different phases in the solar activity cycle. Therein, solar wind types are identified according to two solar-wind classification schemes based on proton plasma properties.
}
{
Within the limits defined by the measurement uncertainties, the proton density and proton temperature can be reconstructed well. Each parameter was evaluated with multiple criteria. Overall proton speed was the parameter with the most accurate reconstruction, while the oxygen charge-state ratio and magnetic field strength were most difficult to recover. We also analysed the results for different solar wind types separately and found that the reconstruction is most difficult for solar wind streams preceding and following stream interfaces.
}
{
For all considered solar wind parameters, but in particular the proton density, proton temperature, and the oxygen charge-state ratio, parameter reconstruction is hindered by measurement uncertainties. The proton speed, while being one of the easiest to measure, also seems to carry the highest degree of redundancy with the combination of the four other solar wind parameters. Nevertheless, the reconstruction accuracy for the proton speed is limited by the large measurement uncertainties on the respective input parameters. The reconstruction accuracy of sector reversal plasma is noticeably lower than that of streamer belt or coronal hole plasma. We suspect that this is a result of the effect of stream interaction regions, which strongly influence the proton plasma properties and are typically assigned to sector reversal plasma. The fact that the oxygen charge-state ratio ---a non-transport-affected property--- is difficult to reconstruct may imply that recovering source-specific information from the transport-affected proton plasma properties is challenging. This underlines the importance of measuring the heavy ion charge-state composition.}


   \maketitle
%

\section{Introduction}
The properties of the solar wind mostly depend on two factors, the
conditions in the solar source region and the transport history of 
the solar wind until it is measured at a spacecraft. The charge states observed
in the solar wind are determined by the electron temperature in the
solar region and are in good approximation frozen in after the solar wind
leaves the hot corona \citep{geiss1995,aellig1997}. The initial
properties bulk proton speed, proton density, and proton temperature also vary with the solar source region of the
solar wind, but are, in addition, affected by transport
effects.

In the context of our study, we consider the following aspects as transport effects: expansion, wave-particle interactions, collisions, and compression regions as found in stream interaction regions (SIRs). Except for expansion, each of them affect the reconstruction of solar wind properties in our study.
As a result of expansion, the magnetic field, the proton density, and
proton temperature all decrease with increasing solar distance
\citep{marsch1982solar,perrone2019radial}. While expansion affects all
the solar wind at 1 astronomical unit (AU) in the same way, other transport effects impact different types of
solar wind differently.
Depending on the solar source region, at least
two types of solar wind are typically distinguished
\citep{steiger2000composition,zhao2009global,zhao2010comparison,xu2014new,camporeale2017classification}.

Coronal holes have been identified as the source of the (typically) faster
component of the solar wind
\citep{hundhausen1968state,Tu2005,Schwadron2005}, which is associated
with low oxygen charge states, low proton densities, high proton
temperatures, and high magnetic field strength. Coronal hole wind is strongly affected by wave--particle interactions; in particular (real
and apparent) heating of the proton bulk, which explains the high
observed proton temperatures in this solar wind type. Wave--particle
interactions are also assumed to be the cause of the differential
streaming observed in coronal hole wind
\citep{berger-etal-2011,kasper2008hot, kasper2012evolution,
  janitzek2016high}.

An example of the redundant information contained in
the different solar wind parameters is provided by the fact that the high observed proton
temperature in coronal hole wind is an effect of the presence of
Alfv{\'e}n waves in the solar wind \citep{marsch1982wave}. This is
illustrated by \citet{heidrich2020proton}, who show that
explicit information about the magnetic field is not necessary to
identify the same solar wind types as in \citet{xu2014new}.
Our study investigates such redundancies among solar wind parameters.

The properties of the slow solar wind systematically differ from those
of the coronal hole wind. Slow solar wind is typically associated with
high proton densities, low proton temperatures, low magnetic field,
and high (oxygen) charge states
\citep{steiger2000composition,zhao2009global,zhao2010comparison,xu2014new}. These
properties correspond to the properties of closed-field-line regions
on the Sun. However, the exact source regions of slow solar wind and the
corresponding release mechanisms are still a matter of debate
\citep{Schwadron2005,Sakao2007,rouillard2010intermittent,Antiochos2011,stakhiv2015origin,
  d2015origin}. At 1 AU, the slow solar wind has experienced just enough collisions that their impact begins to thermalise the velocity distribution function \citep{kasper2012evolution,janitzek2016high} and Alfv{\'e}nic wave activity is low. Solar wind originating in equatorial coronal holes can also
be observed with comparatively low solar wind proton speeds. Such slow
coronal hole wind is also called Alfv{\'e}nic slow solar
wind
\citep{d2015origin,panasenco2020exploring,louarn2021multiscale}. Observing both coronal hole wind and slow solar
wind in the same proton speed range also illustrates that the solar wind proton speed
alone is not well suited to characterizing solar wind. Other solar wind
properties are better tracers of solar wind type.  As observations
from Helios \citep{marsch1982wave}, Parker Solar Probe
\citep{verniero2020parker,zhao2021mhd}, and Solar Orbiter
\citep{jannet2021measurement,carbone2021statistical} show that
waves occur frequently in all types of solar wind close to the Sun,
the presence or absence of waves can also be considered as a transport
effect that is more important close to the Sun than at greater solar
distances.

Another important transport effect that is increasingly
influential  as the solar wind travels further  from the Sun is linked to the compression regions in SIRs that develop at the
boundary of solar wind streams with different speeds
\citep{smith1976observations,richardson2018solar}. If a faster solar
wind stream interacts with a preceding slower solar wind stream, an SIR
forms that is characterised by (hot and dense) compression regions in both the slow
and the fast participating solar wind stream and a high magnetic field
strength at the stream interface. As modelled in
\citet{hofmeister2022area}, SIRs evolve with radial distance and a decreasing amount of unperturbed fast solar wind is observed with increasing
distance. Therefore, in this study, we consider SIRs as a transport effect on the solar wind. Since SIRs are often associated with a
change in magnetic field polarity, in the \citet{xu2014new}
categorisation, compressed slow solar wind tends to be identified as
so-called sector reversal plasma \citep{heidrich2020proton}. 

Although the properties of slow solar wind can be highly variable and
coronal hole wind also shows variability \citep{zhao2014polar,heidrich2016observations} on multiple scales, the respective average
properties are systematically correlated with each other \citep{lepri2013solar,mccomas2000solar,vonSteiger2000}. This
redundancy hints at a common underlying cause that determines these
properties. Under the assumption that all observed solar wind
parameters are determined by the same set of hidden variables in the
solar corona, it would be possible to reproduce each solar wind
parameter from the redundant measurements of the other solar wind
parameters.

In this study, we test this assumption with the help of a
general function approximator to model the (partly) unknown dependencies of the
respective solar wind properties. After the solar wind leaves the
solar corona, such a relation can be modified by transport
effects. Therefore, we investigate the resulting reconstruction
separately for different solar wind types with their different respective
transport histories. In this way, our study evaluates the degree to which the relationship between solar wind parameters is modified by different transport effects. To this end, we employ feed-forward neural networks as general
function approximators \citep{hornik1989multilayer} and apply our method to solar wind observed at
L1. In recent years, the application of machine learning  to solar physics
questions has become increasingly popular. For example, unsupervised
clustering methods are very well suited to solar wind classification
\citep{heidrich2018solar,amaya2020visualizing}. \citet{camporeale2017classification} provide a generalisation
of the \citet{xu2014new} method, with a supervised learning approach based on
Gaussian processes. Ambitious projects aim to predict the solar wind
speed directly from remote sensing observations of the solar corona
with deep neural network architectures \citep{upendran2020solar, raju2021cnn}.
Simple neural networks have been successfully applied as general
function approximators in many different research areas
(e.g. \citet{kuschewski1993application,an1993reservoir,smits1994using,heidrich2009neuroevolution,tahmasebi2011application}) and are therefore well suited to our purposes.

The main goal of our study is to investigate how the relationship between the considered solar wind parameters depends on transport effects. To this end, we compare how accurately each solar wind parameter can be reconstructed from the others under different solar wind conditions, with different dominant transport effects. The relationship between different solar wind properties depends on the solar source region. All effects that further modify this relationship after the solar wind leaves the Sun are considered as transport effects in this study. This includes an increase in the proton temperature due to wave--particle interactions, a systematic increase in the proton speed $\vp$ and the proton temperature $\Tp$ derived from moments of proton velocity distributions that contain a beam, and increased proton density, proton temperature, and magnetic field strength in compression regions in SIRs. The importance of these transport effects is different for different solar wind types: wave--particle interactions are most important in coronal hole wind; collisions become more relevant as the solar wind slows and becomes more dense (and therefore affect slow solar wind), and compression regions are typically found in sector reversal plasma associated with SIRs. In addition, by investigating the impact of measurement uncertainties on our results, our approach provides guidelines as to which solar wind parameters need to be measured with high accuracy.
 
There are several semi-empiric models of the solar wind \citep{arge2000improvement,cranmer2005,cranmer2007self,van2010data, pizzo2011wang, schultz2011space,van2014alfven,pomoell2018euhforia} that derive the solar wind properties
at arbitrary positions in the heliosphere through magneto-hydrodynamic
(MHD) simulations based on observations of the solar photosphere or
the source surface. This is a challenging task, particularly because the release
mechanisms of slow solar wind are still unknown and it is not obvious
whether the observations that provide the boundary conditions for these simulations contain
all the underlying relevant properties of the solar corona at sufficient
resolution. Nevertheless, these models manage to derive the properties
of pure slow and coronal hole wind streams with reasonable accuracy. However, SIRs tend
to be modelled less accurately.
Our approach serves as a minimal sanity check for these kinds of models in two respects: First of all, we can determine whether or not all of the considered solar wind properties are determined
by the same set of (unknown) properties in the solar corona. Second, we attempt to determine the degree to which transport effects obscure a potentially underlying
relationship between different solar wind parameters. 

In addition, our approach can also be applied to alleviate the problem of data
gaps in solar wind data sets in cases where only some but not all
quantities are available. As the solar wind properties of interest
are determined by different instruments, such situations occur
repeatedly because the corresponding data gaps usually do not line up.
Of particular interest is the question of whether charge-state ratios, such as the oxygen $\Oseven$ to $\Osix$ ratio, can be reproduced
from the measurements of proton plasma properties and the magnetic field strength alone. On the one hand,
this would imply that a property that is not affected by transport
effects but is solely determined by the solar origin can be recovered
from the plasma properties that are (strongly) affected by transport
effects. On the other hand, this could help with situations where  information on the
charge-state of heavy ions is not available. Measuring the
charge-state composition of the solar wind is a challenging task and
the resulting instruments have repeatedly suffered from
difficulties. Therefore, for many points in time and space within the
heliosphere, only observations of the proton plasma properties  are available but no
charge-state measurements. If charge-state
information could be recovered (even with low accuracy), this could be
employed to augment existing data sets.

Our neural network approach to reconstruct solar wind parameters is
described in detail in Sect.~\ref{sec:method}. This includes the
preprocessing applied to the solar wind data from the Advanced
Composition Explorer (ACE). In Sect.~\ref{sec:results} we present
and analyse the results of this reconstruction. Our results are
discussed in Sect.~\ref{sec:conclusion}.

\section{Data and methods}\label{sec:method}
We use solar wind data from the Advanced Composition Explorer (ACE) measured by the Solar Wind Electron Proton And Alpha Monitor (SWEPAM, \citet[]{mccomas1998solar}), the magnetometer (MAG, \citet[]{smith1998ace}), and the Solar Wind Ion Composition Spectrometer (SWICS, \citet[]{gloeckler-etal-1998}) from 2001-2010. All data products are binned to the native 12 minute time resolution of SWICS and the only data points considered are those that contain valid entries for proton speed $\vp$, proton density $\np$, proton temperature $\Tp$ (from SWEPAM), the magnetic field strength $\B$ (from MAG), and the oxygen charge-state ratio, $\nO$, with $n_{O^{6+}}$, $n_{O^{7+}}$ as the $O^{6+}$ and $O^{7}$ densities measured by SWICS. Each 12 minute bin is treated as its own isolated data point. Thus, our method does not exploit or model the temporal structure of the solar wind. We assume a time-independent relationship. We test the limits of this assumption by analysing the dependency of the results of our approach in Sect.~\ref{sec:solarcycle}. The data set used in this study is available at \citet[]{Berger2023-gd}.

We chose the 12 min SWICS time resolution as a compromise: it is sufficiently short that we are able to catch short-term variations, while being  as long as is necessary to be able to include charge-state composition data. O$^{6+}$ is the most abundant ion (heaver than He) that is measured in SWICS. Although O$^{7+}$ is less abundant, $\nO$ is among the best determined quantities from SWICS (together with Fe, which is instrumentally well separated from other similarly abundant ions). This choice was influenced by the observations that the majority of the 12 minute $\nO$ data points are within reasonable error margins. This can be seen in Figure 5, where the median of the $\chi^2_{\textrm{red}}$ error is just below 1. The Monte Carlo simulations ---which estimate the effect of the measurement uncertainty on the neural network reconstruction--- take the counting statistics of O into account and show that the neural network reconstruction is stable against the sometimes large uncertainties in the oxygen charge-state composition.

\begin{figure*}[]
     \includegraphics[width=\textwidth]{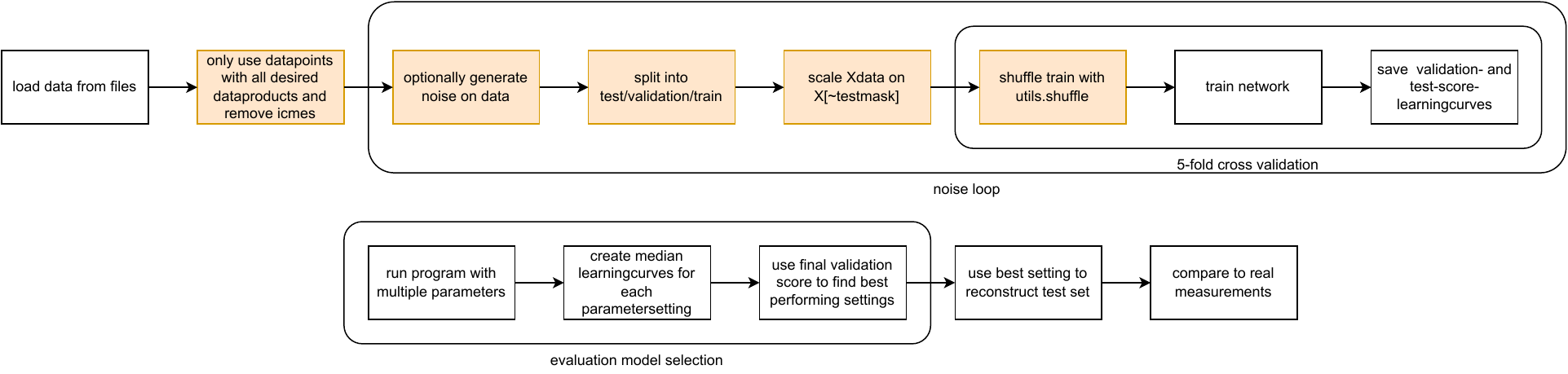}
     \caption{\label{fig:flowchart} Workflow of the solar wind parameter reconstruction algorithm. The top shows the steps done during a single test with specific hyperparameters. The data preprocessing steps are indicated with an orange background.
     The bottom part shows the model selection phase. Here, different hyperparameters are compared and the best hyperparameter combination is chosen to reconstruct the measurements.}
\end{figure*}

In the following, we select four of the five aforementioned solar wind parameters as input parameters for a general purpose function approximator and use this function approximator to reconstruct the remaining fifth parameter. As a general function approximator, we employ a simple feed-forward neural network, namely a multi-layer perception (MLP). This type of neural network is described in more detail in Sect.~\ref{sec:mlp}. Our objective is formulated as a supervised regression task, that is, the neural network is used to model a functional relationship between input and output data and is provided correct output data examples during training. Our experimental setup is described in the remainder of this section and is summarised in Fig.~\ref{fig:flowchart} and the source code is available at \citet[]{Hecht2023}.

\subsection{Preprocessing: data selection}
Before the data are presented to the neural network, we apply the following preprocessing to the ACE data set.  We apply a decadic logarithm to $\nO$. An output variable $\vec{y}_{\text{rec}}\in [ \vp, \np, \Tp, \B, \lnO]$ is then selected for each training scenario. Depending on the chosen output variable $\vec{y}_{rec}$, we construct an input vector $\vec{X}$ from the remaining four solar wind parameters. The output variable $\vec{y}_{rec}$ is the data product that is going to be reconstructed, while the input vector $\vec{X}$ contains the measurements provided for the reconstruction.  

To categorise solar wind types ---and thereby implicitly select solar wind observations with different transport histories--- we employ the scheme presented in \citet[]{xu2014new} or order the data according to proton--proton collisional age, which allows (as shown in \citet{heidrich2020proton}) a very similar solar wind classification.
The proton--proton collisional age is calculated by
\begin{align}
         \ac = \frac{6.4\cdot10^8\text{K}^{3/2}}{\text{cm}^{-3}} \frac{\np}{\vp\Tp^{3/2}} \enspace.
\end{align}
The \citet{xu2014new} solar wind classification scheme distinguishes between coronal hole wind, two types of slow solar wind, and ejecta. The two types of slow solar wind were defined to distinguish between helmet-streamer and pseudo-streamer plasma. Sector reversal (or helmet-streamer) plasma includes a change in the magnetic field polarity and therefore consists of slow and dense slow solar wind in the vicinity of stream interaction regions. Streamer belt (or pseudo-streamer) plasma contains the remaining slow solar wind plasma. The fourth category from the \citet[]{xu2014new} scheme, namely the ejecta category,  which is designed to detect interplanetary coronal mass ejections (ICMEs), is disregarded here because it tends to misidentify particularly cold and dense slow solar wind \citep{sanchez2016very} as ejecta. As ICMEs undergo a (most likely) very different release mechanism from the ubiquitous solar wind, we cannot expect the same relations that hold between  properties in the solar wind to also hold between properties in ICMEs. Therefore, we do not consider ICMEs in the following analysis. Instead, ICMEs are identified based on the ICME list from \citet{cane2003interplanetary,richardson2010near} and \citet{jian2006properties,jian2011comparing} and are subsequently removed from the data set. As the start and end times in both ICME lists are not necessarily well defined, we extended each ICME time interval by six hours at the beginning and the end of each ICME.

\subsection{Test, training, and validation data sets}
To apply and evaluate a supervised learning method, we need to separate the available data set into three different subsets: training, validation, and test data. The training data are used in the training of the neural network, and the validation data are used to estimate the generalisation error of the trained model and for the selection of optimal hyperparameters of the model (see Section \ref{sec:modelsel}). The previously unseen test data set is only used to evaluate the final performance of the model and is the only data set suitable for a comparison between different models.

Therefore, we partition the ACE data set into batches with the approximate length of one Carrington rotation, which is $27.24$d. Now, these batches are split into test, training, and validation data sets. The selection of training, validation, and test data sets is randomised, but to ensure that each data set is well distributed over time, we apply the following to the five two-year time frames in our data set. Herein, for each two-year time frame, we randomly select four data batches of Carrington rotation length as test data. From the remaining ten data batches of each two-year time frame, we select training and validation data sets for a five-fold cross-validation \citep{allen1974relationship,stone1974cross,stone1977asymptotic}, that is, two batches become part of the validation data set and the remaining eight go into the training data set. Five-fold cross-validation helps to improve the generalisation capabilities of supervised learning methods by permuting the role of each fold as a validation data set (while always training on the remaining data). This reduces the dependency of the result on a particular choice of the training or validation data set. 

Next, the \texttt{scikit-learn} \citep{scikit-learn} \texttt{StandardScaler} is applied to the input vector $\vec{X}$ to ensure that the learning is not inhibited by ill-conditioned data points. This standardises the numerical value of each input dimension by removing the mean and scaling to unit variance.
Afterwards, the training data are shuffled. As we consider each 12 minute bin independently, we neglect the temporal information. Randomising the order of data points is beneficial for the learning speed of a neural network. 

Then, after training, each trained neural network is applied to the corresponding validation data set from the five-fold cross-validation. The resulting validation score (see Sect.~\ref{sec:scores}) is averaged over the five folds. The average validation score is then used for model selection (see Sect.~\ref{sec:modelsel}).
The complete experimental framework is illustrated in Fig~\ref{fig:flowchart}.

\begin{table*}[]
\begin{tabularx}{\textwidth}{c|l|XXXXX}\small
hyperparameter      & tested hyperparameter                                         & $\B$  & $\nO$  & $\np$  & $\Tp$ & $\vp$ \\ \hline

$n_h$               & {[}10, 20, 50, 100{]}                                        & 10    & 10     & 10     & 10    & 10    \\
iterations          & 200                                                          & 200   & 200    & 200    & 200   & 200   \\
activation function & relu                                                         & relu  & relu   & relu   & relu  & relu  \\
logarithmic $\nO$   & True                                                         & True  & True   & True   & True  & True  \\
solver              & adam                                                         & adam  & adam   & adam   & adam  & adam  \\
$\lambda$           & {[}0.01, 0.001, 0.0001{]}                                                                  & 0.001  & 0.001 & 0.001 & 0.01  & 0.001  \\
$\beta_1$           & {[}0.75, 0.8, 0.85, 0.9, 0.95, 0.99{]}                                             & 0.99  & 0.9    & 0.75   & 0.9   & 0.95   \\
$\beta_2$           & {[}0.8, 0.85, 0.9, 0.95, 0.99, 0.999{]}                                            & 0.95  & 0.999  & 0.9   & 0.999  & 0.999 \\
$\epsilon$          & {[}$10^{-6}$, $10^{-7}$, $10^{-8}$, $10^{-9}$, $10^{-10}${]}         & 1e-09 & 1e-09  & 1e-08  & 1e-06 & 1e-06 \\
$\alpha$            & {[}0.001, 0.0001, 0.00001{]}                                                               & 0.0001 & 1e-05  & 0.001  & 0.0001 & 0.001
\end{tabularx}
\caption{Investigated (column 2) and best-performing (columns 3-7) hyperparameters from the model selection for each solar wind parameter. The best hyperparameters are used in the final model as well as in the Monte Carlo error simulations.}
\label{tab:used hyperparameter}
\end{table*}

\begin{figure}[]
     \includegraphics[width=.7\columnwidth]{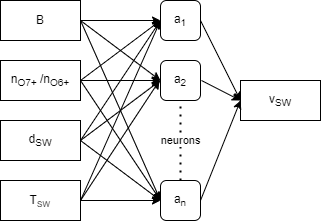}
     \caption{\label{fig:neural net} Schematic overview of our neural network architecture. Four solar wind parameters form the input vector $\vec{X}$ (in this example $\B$, $\lnO$, $\np$, and $\Tp$), the hidden layer contains a variable number of neurons $n_h$, and the output $\vec{y}_{\text{rec}}$ is the remaining solar wind parameter (in this example $\vp$). Each layer is fully connected by weights that are represented by arrows.}
\end{figure}  

\subsection{Multilayer perceptron }\label{sec:mlp}
We employed a multilayer perceptron (MLP) with one hidden layer as a general purpose function approximator \citep{hornik1989multilayer} to predict a data product $\vec{y}_{\text{rec}}$. The input vector $\vec{X}$ consists of the remaining four data products. The specific implementation used is the \texttt{MLPregressor} from the Python package scikit-learn \citep{scikit-learn} version 1.1.2.

Figure~\ref{fig:neural net} shows a schematic overview of our neural network structure. In addition to the input vector $\vec{X}$ and the output, which is the reconstructed value $\vec{y}_{\text{rec}}$, the setup includes a hidden layer consisting of $n_h$ neurons. Each layer is fully connected to the next layer by weights $w_{i,j}\not=0\in \mathbb{R}$ (with $i,j$ indices of neurons from two consecutive layers, e.g. input to hidden layer or hidden layer to output). These form a weight matrix $\textbf{W}$. The general principle of neural network training can be summarised in three steps: (1) The forward pass. For each layer, the product of the input vector (or neuron vector) with the weight matrix is computed and the activation function $f$ is applied to obtain the input vector for the next layer $\vec{h}$ or the result $\vec{y}_{\text{rec}}$: $\vec{h}=f(\textbf{W}\vec{X})$. (2) Computation of the difference between known output and current output. The difference between the calculated value $\vec{y}_\text{rec}$ and the measured value $\vec{y}_\text{meas}$ describes the current training progress and is calculated using the mean squared error. (3) Back-propagation. The aforementioned difference, as the estimated training progress, is minimised by propagating the error information from the output layer backwards through the neural network. This changes the values in the weight matrices and minimises the training error. Here, we employ the efficient Adam solver; see \citet{kingma2014adam} for a full description. These three steps are repeated $n_{\text{iter}}$ times. 

  \begin{table}
\centering\small
\begin{tabularx}{\columnwidth}{l|lllll}
parameter  & $\vp$ & $\np$ & $\Tp$ & $B$ & $nO$  \\\hline
$\Delta$ & 1.5\%   & 15\%     & 20\% & 0.1 nT  & \text{from counts} 
\end{tabularx}
\caption{\label{tab:noise} Relative measurements errors $\Delta$ of solar wind parameters according to \cite{smith1998ace,skoug2004extremely, berger2008velocity}.}
\end{table}

\begin{figure}[]
     \includegraphics[width=\columnwidth]{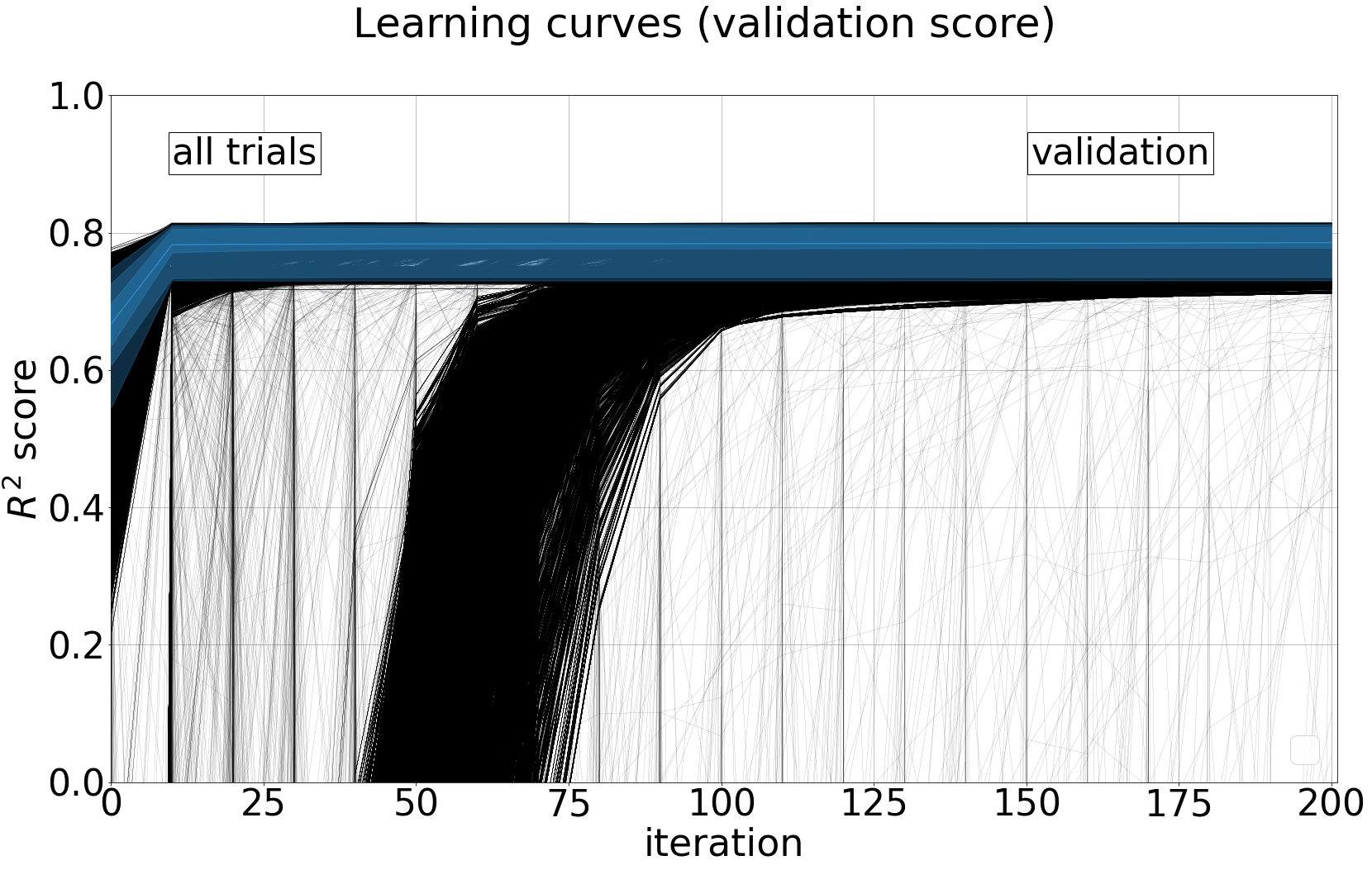}
     \caption{\label{fig:valid} Validation scores for all considered hyperparameter configurations with ten neurons and all 10 trials for $\vec{y}_{\text{rec}}=\vp$. Each individual trial is shown with a thin black line. The hyperparameter combination with the highest final median validation score is plotted in blue. In addition, the median and three overlapping confidence intervals (15.9th - 84.1th percentilel, 2.5th - 97.5th percentiles, and 0th - 100th percentiles) range are shown in overlapping blue shaded regions.}
\end{figure}

As the chosen back-propagation variant, Adam, has stochastic elements, we repeated the training for 100 independent trials. Each trial is initialised with a different random seed for the initial weights and the training data is shuffled for each trial. Training is stopped after 200 iterations. As shown in Fig.~\ref{fig:valid}, the performance appears to converge after less than 200 iterations for the majority of the hyperparameter combinations. While there is no guarantee that more iterations will not yield further improvements, some tests with 2000 iterations showed no indications of this.

\subsection{Error estimation and stability}\label{sec:scores}
In this subsection, we describe the different performance and error estimates that are of interest for our study.
First, to evaluate the validation error, which is the basis for selecting optimal hyperparameter settings, we employ the $R^2$ score. For each reconstructed solar wind parameter, the reconstruction $\vec{y}_\text{rec}$ is compared to the measurements $\vec{y}_\text{meas}$. The score $R^2$ is calculated by
\begin{align}
        R^2 &= \left ( 1-\frac{r}{s} \right ) \label{eq:R2} \enspace ,\\
        r &= \sum^m_{i=1} (y_{\text{meas},i} - y_{\text{rec},i})^2 \enspace ,\\
        s &= \sum^m_{i=1} (y_{\text{meas},i} - \text{mean}(y_{\text{meas},i}))^2 \enspace .
\end{align}
An $R^2$ score of 1  indicates a perfect reconstruction. The $R^2$ score is used in the model selection to choose the specific hyperparameters for each reconstruction. The $R^2$ score is well suited to comparing different hyperparameter configurations for the same reconstructed parameter and on the same data set.
For a comparison of each of the reconstructed solar wind parameters, different neural networks have to be assessed in relation to each other. For this, the $R^2$ score, which depends on the estimated variance of the data set, is less well suited. 

The results of our study are affected by different sources of uncertainty. Therefore, in the following, three different types of error or uncertainty measure are considered. The first type of uncertainty is the measurement error $\Delta$ from the measurements of the solar wind parameters. For $\Delta\vp, \Delta\np,$ and $\Delta\Tp,$ relative errors are taken from the literature \citep{skoug2004extremely}. For $\Delta\B,$ an absolute value of $0.1$ nT is given by \citet{smith1998ace}. For $\Delta\lnO$, the error is derived from the actual counting statistics of SWICS based on Poisson statistics. As $O^{7+}$ is rare and can be at the limit of the detection capabilities of SWICS in very dilute solar wind, the resulting error can be enormous. We decided against excluding the data points with particularly high oxygen charge-state measurement errors because these occur systematically in very dilute coronal hole wind mainly during the solar activity minimum. Therefore, excluding these data points from our analysis would introduce a systematic bias in the data set. These errors and the reference they were taken from can be seen in Table \ref{tab:noise}.

The second type of error is defined by the comparison of the original measured data to the reconstructed values. These errors are only calculated on the test data set, which encompasses a sample size of 63432 points. For this purpose, we consider linear and quadratic measures. A first straightforward approach is to calculate the relative reconstruction errors $y_\text{diff,p}$ between the observed and reconstructed quantity: 
\begin{equation}
        y_{\text{diff}} = \frac{y_\text{meas} - y_\text{rec}}{y_\text{meas}} \enspace .
\end{equation}
Further insights are provided by the mean absolute percentage error (MAPE) and the reduced $\chi^2_{\text{red}}$ score. These errors are used to evaluate the accuracy of the reconstruction or as a goodness of fit parameter. The MAPE is calculated as
\begin{align}
         \text{MAPE} &= \frac{1}{m}\sum^n_{i=1} \left | \frac{y_{\text{rec},i} - y_{\text{meas},i}}{y_{\text{meas},i}} \right | \enspace ,\label{eq:MAPE}
\end{align}
with $m$ being the number of samples and $i$ the index of each sample. An approximation of the reduced chi-square statistic is calculated on the test data set:
\begin{align}
        \chi^2 &= \sum_i \frac{(y_\text{meas, i} - y_\text{rec, i})^{2}}{(\Delta y_\text{meas, i})^{2}}, \label{eq:chi2_sum} \enspace \\
        \chi^2_\text{red} &= \frac{\chi^2}{\nu} \enspace ,
\end{align}
with the degrees of freedom $\nu$, which is here given as the sample size of the test set 
(63432) minus the number of parameters fixed by the model, that is, the number of connections in the neural network  (here 61).

Typically, the reduced $\chi^2_{\text{red}}$ score is used in the context of fitting and is computed for all summation indices $i$ in Equation \ref{eq:chi2_sum} in the data set to which the model was fitted. Here, we apply this concept to evaluate the reconstruction error with respect to the measurement uncertainty. Therefore, in our case, the reduced $chi^2_{\text{red}}$ is calculated from the test data set (and not the training data).  The $\chi^2$ score is still considered as a measure of goodness of fit and augments the comparison between the five reconstructions based on the MAPE score. In the context of fitting, a $\chi^2_\text{red}$ of one indicates a good fit consistent with the measurement errors. Lower values of the reduced $\chi^2_{\text{red}}$ can indicate overfitting due to large measurement uncertainties.

The third type of error in our study is an estimate of the impact of the measurement uncertainties on the reconstruction error. This estimate is derived with a Monte Carlo simulation and the resulting error is therefore also called the Monte Carlo error.  As the potential accuracy of the reconstruction by the MLP regressor is limited by the underlying measurement uncertainty of the five considered solar wind parameters, a basic Monte Carlo approach is used to estimate the effect of this measurement uncertainty. To this end, Gaussian noise is added to each data point. The respective standard deviations of these Gaussian noise distributions depend on the data products  to be reconstructed and their respective  measurement errors $\Delta,$ and are listed in Table~\ref{tab:noise}. For each noisy data set generated in this way, we apply the same procedure as described in the previous subsections. We repeat the Monte Carlo simulation 100 times. The distribution of the resulting reconstructions based on the noisy data set provides a measure of the susceptibility of the MLP regressor to the measurement uncertainty. To ensure that the Monte Carlo simulation is not biased by the occasionally very poor statistics of $\Osix$, we limit the Monte Carlo noise of $\lnO$ to 0.41 and use this value for the 14.9\% of the oxygen data that exhibit a larger relative measurement uncertainty. The variability of the Monte Carlo results is indicated by confidence intervals corresponding to a $1\sigma$ environment defined by the $15.9$th and $84.1$st percentiles. We refer to these as 1$\sigma$ equivalent percentile confidence intervals in the following. 


\begin{figure*}[]
  \includegraphics[width=\textwidth]{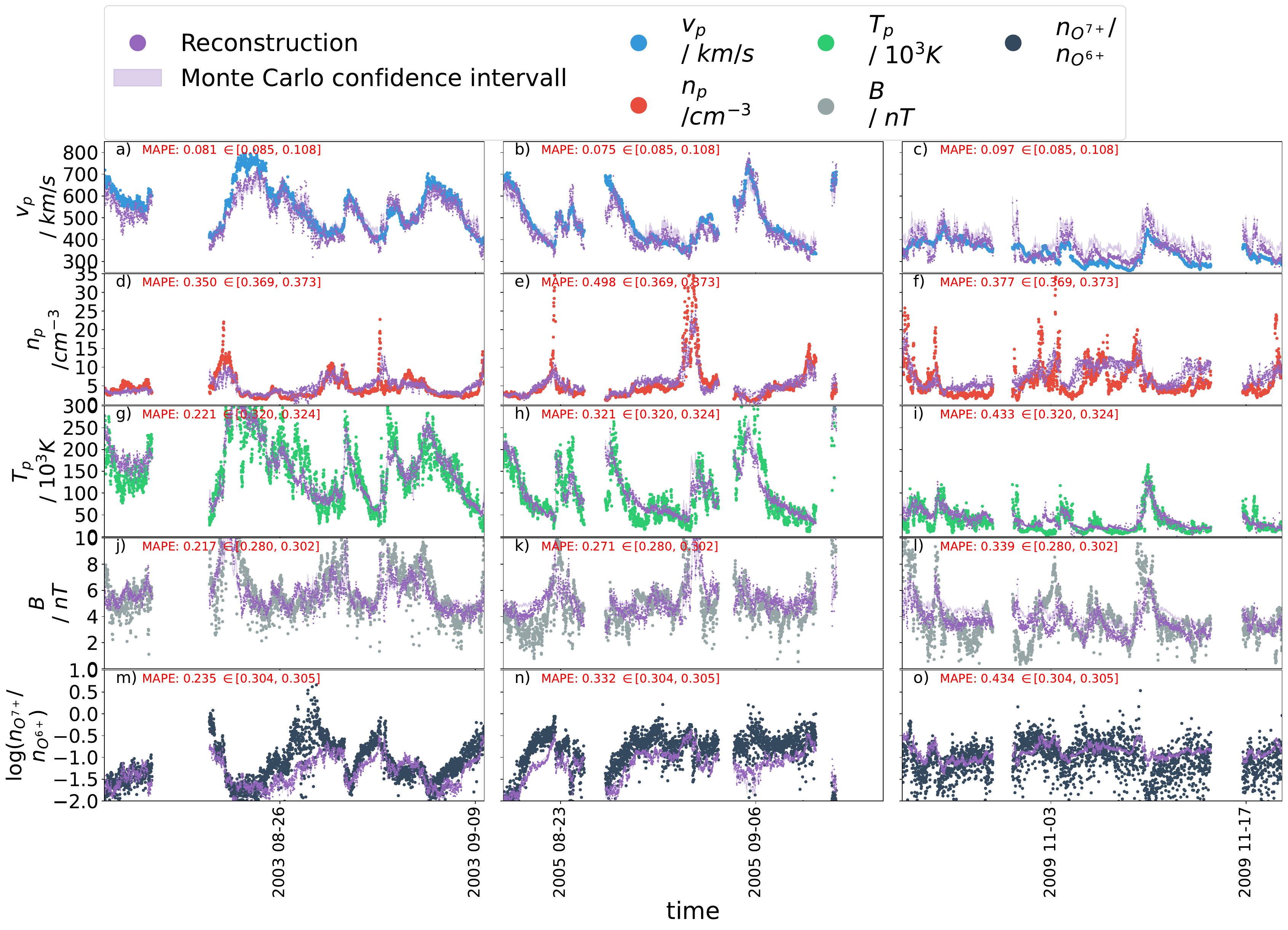}
  \caption{\label{fig:test_pred} Time series of reconstructed and measured solar wind parameters for three selected test data time periods with the average length of a Carrington rotation. The time periods were chosen, based on the MAPE score, to be representative examples of `good', `intermediate', and `poor' performance, respectively. The time period to the left represents one of the most accurate reconstructions, the middle time period represents a reconstruction of average accuracy, and the right time period is one of the poorest reconstructions. For each Carrington rotation, the observed data are plotted in blue and the reconstructed data are plotted in purple. The uncertainty on the reconstruction is estimated with 100 Monte Carlo simulations. The 15.9th to the 84.1st percentiles of the Monte Carlo runs are plotted as purple-shaded areas. Each row depicts one solar wind parameter, from top to bottom: $\vp$ (a), (b), and (c) in blue, $\np$ (d), (e), and (f) in red, $\Tp$ (g), (h), and (i) in green, $\B$ (j), (k), and (l) in light grey, and $\nO$ (m), (n), and (o) in dark grey. The MAPE score for each part of the test data set is shown in red as an inset.}
\end{figure*}

\subsection{Model selection}\label{sec:modelsel}
The performance of a machine learning method depends (often sensitively) on the choice of hyperparameters of the method. Therefore, unbiased evaluations and comparisons of machine learning methods are only feasible if optimal hyperparameters are used. The process of selecting hyperparameters is called model selection. Here, we employ a simple grid search to select optimal hyperparameters. Table \ref{tab:used hyperparameter} summarises the hyperparameters of the MLP regressor in \texttt{sci-kit learn} and our overall method (see also Fig.~\ref{fig:flowchart}). For each combination of hyperparameters given in Table \ref{tab:used hyperparameter}, we trained our neural network and computed the validation error.

The considered hyperparameters includes the number of neurons $n_h$, the initial learning rate $\lambda$, the L2 penalty $\alpha$, the exponential decay rate for estimates of the first moment vector $\beta_1$ and the second moment vector $\beta_2$, and the value for numerical stability $\epsilon$. We also performed tests with different activation functions, finding that replacing \texttt{relu} with the logistic function or a $\tanh$ does not have a significant impact on the results of our study.

We base the model selection not on the complete learning history as shown in Fig.~\ref{fig:valid} but on the final validation scores after 200 iterations. Due to the high number of hyperparameter combinations tested, we
initially conducted only ten trials for each hyperparameter configuration. The resulting variability is illustrated in Fig.~\ref{fig:valid} where the performance of each individual trial for all combinations of the hyperparameters in Table \ref{tab:used hyperparameter} is shown for $\vec{y}_{\text{rec}}=\vp$. Figure~\ref{fig:valid} illustrates that many (most) hyperparameter combinations lead to very similar final validation scores. The variability of the hyperparameter combination with the highest final median validation error is indicated with the blue shaded area. This shows that the uncertainty from the individual trials is larger than the differences in the median performance of different hyperparameter combinations. For each $\vec{y}_{\text{rec}}$, the hyperparameter combination with the highest final median validation score is considered as optimal. The optimal hyperparameters chosen in this way depend on which solar wind parameter is chosen as the reconstructed output vector $\vec{y}_{rec}$. These optimal combinations are used in the remainder of this study to reconstruct the solar wind parameter on the test set. Table \ref{tab:used hyperparameter} shows the final hyperparameters.

\section{Reconstruction of solar wind parameters}\label{sec:results}

\begin{figure*}[]
     \includegraphics[width=\textwidth]{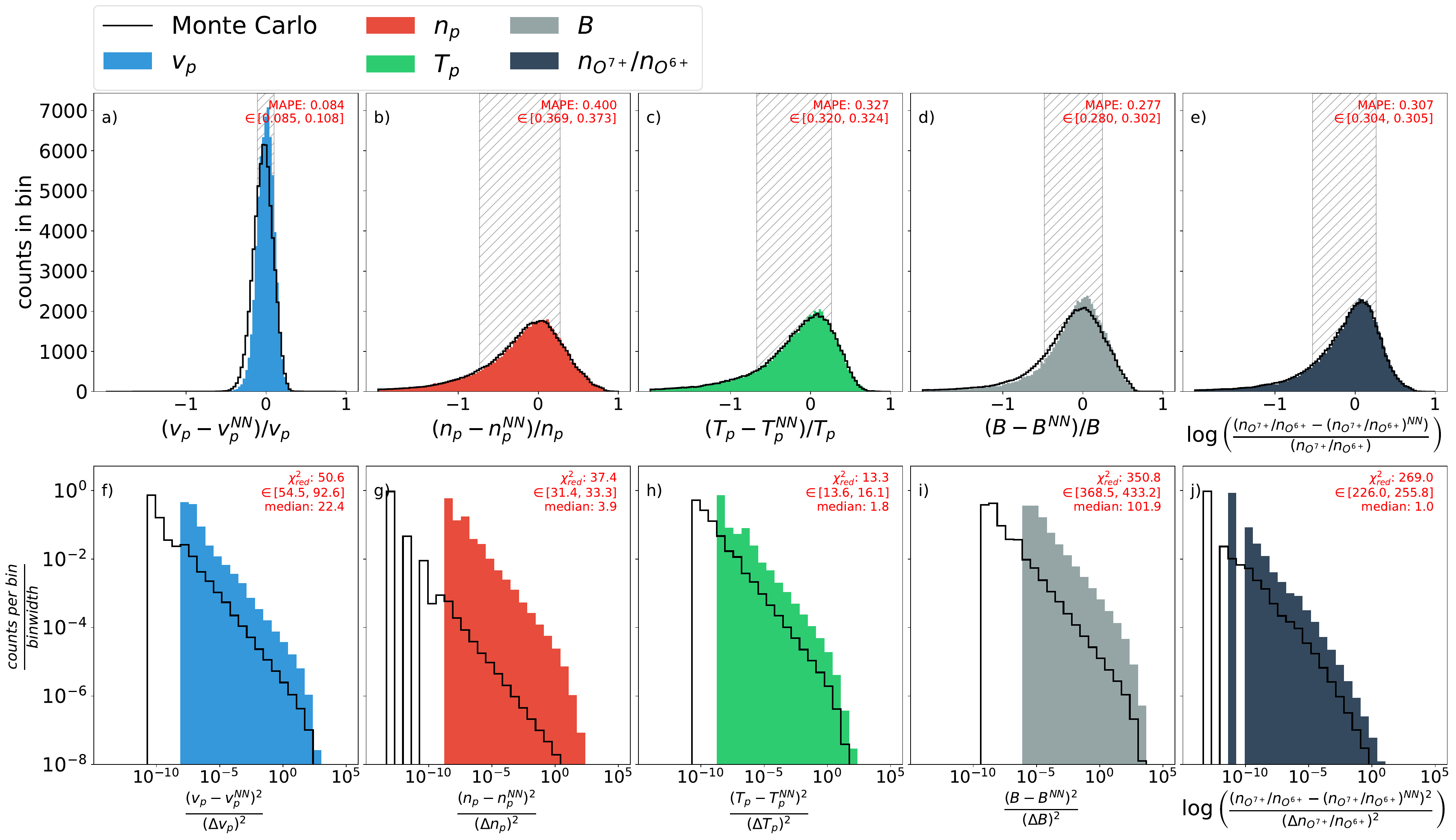}
     \caption{\label{fig:test_hist} One-dimensional histograms of reconstruction errors $y_{\text{diff}}$ and $\chi^2$ scores for each reconstructed solar wind parameter. In each panel, the x-axis is constrained to contain 100 bins from -2 to 1. Each of the top panels depicts the reconstruction accuracy of one of the five reconstructed solar wind parameters (a-e) based on the MAPE score, and the bottom panels (f-j) show normalised densities of the $chi^2$ score for each determined parameter. Each column refers a different reconstructed parameter, from left to right: $\vp$ (a) and (f) in blue, $\np$ (b) and (g) in red, $\Tp$ (c) and (h) in green, $\B$ (d) and (i) in light grey, and $\nO$  (e) and (j) in dark grey. In each histogram, an area is marked with grey hatches that contains all data from the $15.9$th to the $84.1$th percentiles. The respective MAPE score and   $\chi_{\textrm{red}}^2$  are indicated in insets in the top and bottom rows. For both, the respective confidence intervals derived from the Monte Carlo runs are included. An additional black histogram outline is included in each panel, which represents the variability of the Monte Carlo simulations based on randomised input data. In each panel, the x-axis contains 30 logarithmic bins. In the bottom row only, the y-axis is also logarithmic with a lower bound of $10^{-8}$, i.e. 0.00001\% of the $\chi^2$ score density, and the histogram is normalised to the sum of the distribution. The inset also depicts the $\chi^2_\text{red}$ score. In addition, the median of the individual $\chi^2$ scores is noted to show the spread of the distribution.}
\end{figure*}

\begin{figure}%
\includegraphics[width=\columnwidth]{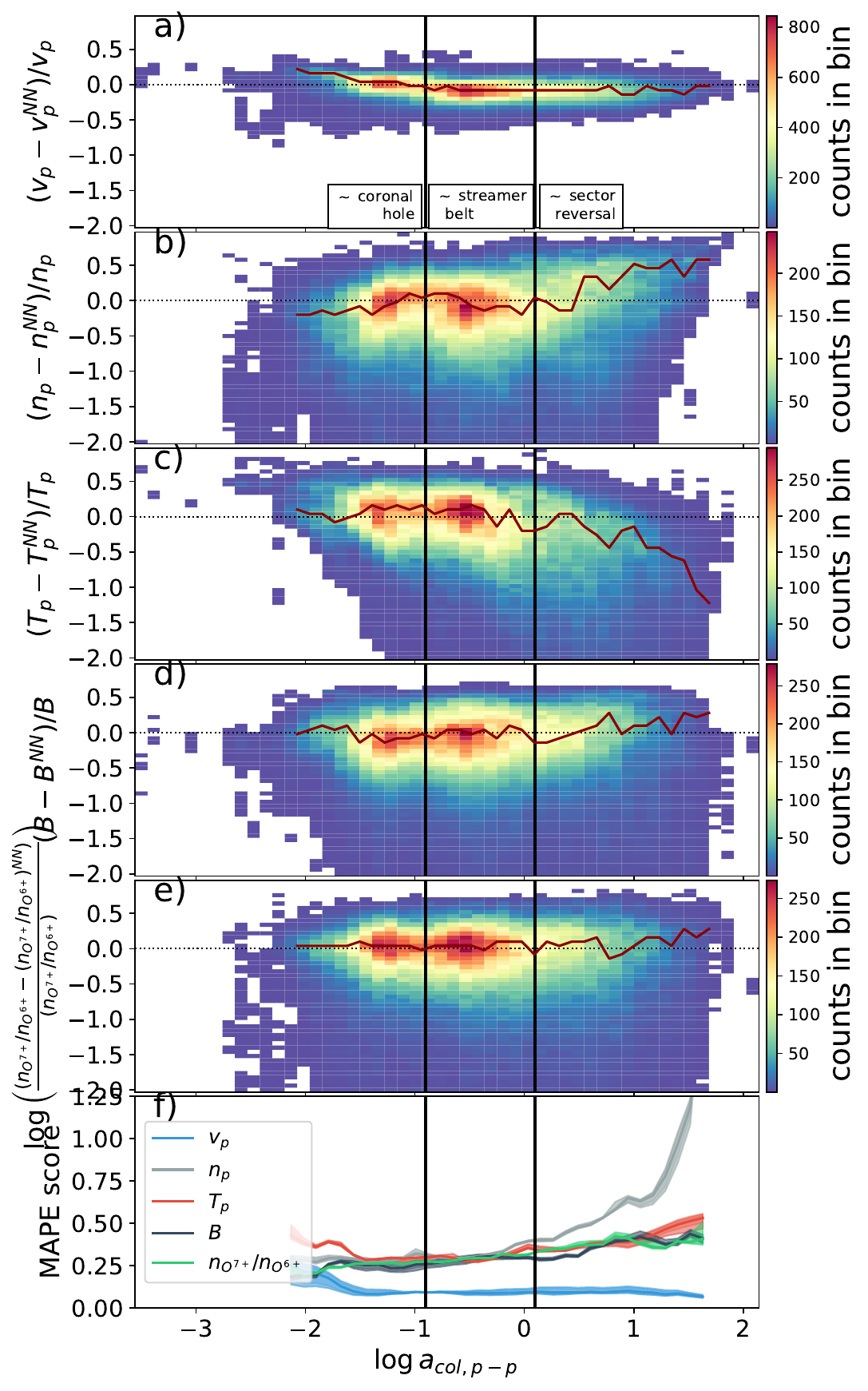}

\caption{
Two-dimensional histograms of the relative reconstruction error (a, b, c, d, e) and MAPE score (f))  for each reconstructed solar wind parameter over the proton--proton collisional age: From top to bottom: $vp$, $\np$, $\Tp$, $\B$, and $\nO$. For the first five subplots, the y-axis shows the normalised differences between the reconstruction and the measured data. The x-axis shows the proton--proton collisional age. In all panels, data are sorted into 50 bins between -3.5 and 2.2 on the x-axis, and in panels (a-e) the data are sorted into 50 bins between -2.0 and 1.0 on the
y-axis. The vertical black lines give approximate thresholds separating sector reversal from streamer belt plasma (right black line) and streamer belt plasma from coronal hole wind (left black line). The red line highlights the maximum of the distribution in each vertical slice with sufficient statistics (at least 5000 data points over all Monte Carlo runs per column). The bottom-most subplot shows the MAPE scores (on the y-axis) computed separately for all test data points falling into the respective proton--proton collisional age bin (on the x-axis) for each reconstructed solar wind parameter. The MAPE score is calculated according to Equation \ref{eq:MAPE}. Confidence intervals in panel (f) are given as three overlapping areas per curve ($15.9$th - $84.1$th percentile, $2.5$th - $97.5$th percentiles, and $0$th - $100$th percentiles).}\label{fig:colage}
\end{figure}

\begin{figure}[]
  \includegraphics[width=\columnwidth]{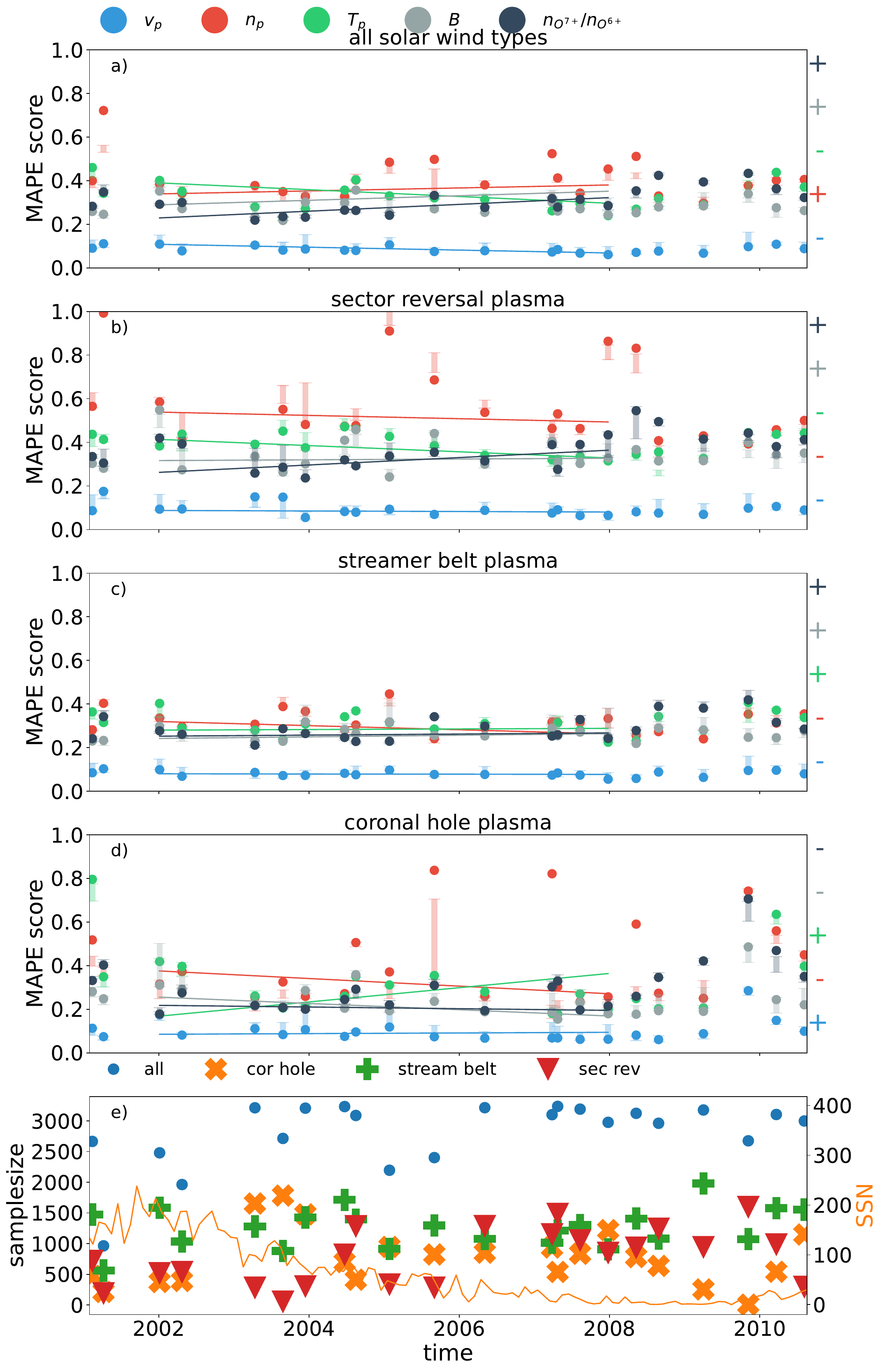}
  \caption{\label{fig:test_scores} MAPE scores for each time period from the test data set for each reconstructed solar wind parameter over ten years and sorted by solar wind type. The top panel shows the respective MAPE scores on all (non-ICME) solar wind data from each test data time period. The three panels below show the MAPE scores separated by their \citet{xu2014new} solar wind type (from top to bottom: sector reversal wind, streamer belt plasma, and coronal hole wind). The bottom panel gives the number of valid data points per test data time period and solar wind type. In addition, on the right y-axis, the bottom panel includes the monthly sun spot number as a reference \citep{sidc}. In the four top panels, simple linear fits to the scores of each reconstructed parameter (from 2002-2008) are shown as thin coloured lines. To the right of each of the four upper plots, plus and minus symbols in the colour of the respective reconstructed parameter indicate whether the slope of the corresponding line is positive ($+$) or negative ($-$).}
\end{figure}
    
We apply our method to the ACE data set as described in the previous section to obtain a model (realised by a neural network) for each of the five considered solar wind parameters. This model can then be applied to previously unseen solar wind data, namely the test data, to evaluate and analyse the performance of our neural network function approximators and to address our research questions.

Figure~\ref{fig:test_pred} shows 3 of the 22 test data time periods of 27.24 days. For each reconstructed parameter, the respective observation is shown in the same panel. As an inset, in red the MAPE score of the reconstruction is given in each panel together with confidence intervals derived from the Monte Carlo simulations. These confidence intervals reflect the effect of the measurement uncertainty estimated by Monte Carlo simulations and are defined by the $15.9$th and $84.1$st percentiles of the Monte Carlo simulation results. As these values are calculated from noisy data, the score of the original data is in some cases significantly different (see in Fig. \ref{fig:test_pred}c)). Overall, the reconstruction captures the major fluctuations in all solar wind parameters reasonably well. However, for all reconstructed solar wind parameters, particular low and high values in the observations are frequently over- or underestimated by the neural network. These observations are supported by the calculated MAPE scores. In particular, $\nO$ is consistently underestimated in the second time period (Fig.~\ref{fig:test_pred}n)) and for most of 2003 August 26 in the first time period (Fig.~\ref{fig:test_pred}m)). Further, the reconstruction quality varies for different reconstructed solar wind parameters and in different test data time periods. In particular, the reconstruction of the proton speed appears more accurate than the other reconstructions. This topic is investigated in more detail in the following subsections. 

\subsection{Reconstruction error}\label{sec:reconstruct}

The upper row of Fig.~\ref{fig:test_hist} shows histograms of relative reconstruction errors for the complete test data set and all reconstructed solar wind parameters. Each panel also gives the MAPE score for each reconstruction. A MAPE score of zero indicates a perfect reconstruction. The sign differentiates between overestimation (negative sign) and underestimation (positive sign). An overestimation of 50\% would result in a MAPE score of -0.5 and an overestimation by 100\%, or double the measured value, would lead to a MAPE score of -1.0. An underestimation by 50\%, or half the measured value, results in a MAPE score of +0.5.  The confidence intervals of the MAPE score were calculated for the 100 Monte Carlo runs and are given as 1$\sigma$ equivalent percentile confidence intervals.
Each histogram in Fig.~\ref{fig:test_hist} is augmented by a black outline that summarises the results of the Monte Carlo simulations.
 
We now first focus on the MAPE scores (included in the legend in each panel). The reconstruction of the proton speed results in a MAPE score of $0.084 \in[0.085,0.108]$. In comparison, the other reconstruction errors are $\np: 0.400 \in[0.369,0.373]$,  $\Tp: 0.327 \in[0.320,0.324]$, $\B: 0.277 \in[0.280,0.302],$ and $\nO: 0.307 \in[0.304,0.305]$. Therefore, as already illustrated in Fig.~\ref{fig:test_pred}, the reconstruction of the proton speed appears more accurate than the other reconstructions. This is also apparent in the shape of the histograms. The histograms of reconstruction errors for the four other solar wind parameters are asymmetric and feature heavier tails biased towards negative values (Fig.~\ref{fig:test_hist} (b), (c), (d), and (e)). This means that the reconstructions for these solar wind parameters tend to overestimate the observations more frequently and more strongly than they underestimate them. The 1$\sigma$ equivalent percentile confidence intervals of the distribution of the individual reconstruction errors (indicated with grey hatching) also underline this. The lower percentile bound is further away from zero than the upper percentile border. In addition, while the maxima of the histogram for the proton speed and the magnetic field strength are located at zero, the maxima are shifted to the right for the three other solar wind parameters. Thus, while extreme values are frequently overestimated by the neural network for $\np$, $\Tp$, and $\nO$, most values are slightly underestimated. This indicates that the model attempts to smooth the solar wind observations more than desired. While this could be a consequence of an excessively small hidden layer in our neural network, our model selection does not indicate an improvement of the validation error for larger hidden layer sizes. Therefore, we favour different explanations: Either (1) the reconstruction might be inhibited by the measurement accuracy, or (2), given that our model is time-independent, within the limitations of the measurement uncertainties, short-term variations that affect some but not all of the considered solar wind parameters cannot be captured by the neural network reconstruction. 

The lower row of Fig.~\ref{fig:test_hist} shows histograms of the individual $\chi^2$ values for each reconstructed parameter. The x-axis shows a logarithmic bin distribution of the $\chi^2_{\text{red}}$ scores with 30 bins. The y-axis shows the density distribution of each $\chi^2_{\text{red}}$ bin. The density distribution is normalised to 1. The red inset provides the reduced $\chi^2_{\text{red}}$ score of the whole data set. The confidence intervals in the second row are calculated by taking the $\chi^2_{\text{red}}$ scores of the Monte Carlo runs and computing the $15.9$th and $84.1$st percentiles. The third line of the inset provides the median value of the individual $\chi^2$ on the test data set (not of the Monte Carlo runs). 

The $\chi^2_{\text{red}}$ scores show that the reconstruction that is closest in line with the measurement errors is the proton temperature $\Tp$ reconstruction with a score of $13.3 \in [13.6, 16.1]$, followed by the proton density $\np$, the proton speed $\vp$, the oxygen charge-state ratio $\nO$, and finally the magnetic field strength $\B$. We interpret the comparatively very poor $\chi^2_{\text{red}}$ score of the proton speed despite the apparently good MAPE scores of the reconstruction as being the result of the small measurement uncertainty on the proton speed (in particular in comparison to the measurement uncertainties and scores of the proton density and proton temperature; see Table \ref{tab:noise}). A similar effect probably impacts the magnetic field strength $\chi^2_\text{red}$ score. For values between 1 and 10 nT, a measurement error of 0.1 nT would correspond to a relative measurement error of 10\% to 1\%. This is a lower relative measurement error than for the proton density $\np$ or the proton temperature $\Tp$. Therefore, the $\chi^2$ score is poorer despite the fact that the reconstruction accuracy estimated by the MAPE score is similar to that of $\np$ and $\Tp$. However, in the case of the magnetic field strength, the reconstruction is also less accurate than that of the proton speed. The charge-state ratio of oxygen is associated with the largest measurement errors (by far) and this is reflected in the poorest $\chi^2_\text{red}$ score.
The median value of the individual $\chi^2$ scores for the proton temperature ($1.8$) and the oxygen charge-state ratio ($1.0$) indicate that the majority of the reconstructed data points are consistent with the measurement errors even though their average, the reduced $\chi^2_{\text{red}}$ , is strongly affected by outliers with a very poor $\chi^2$ score.

\begin{table*}[]
\begin{tabularx}{\textwidth}{c|XX|XX|XX|XX}\small
      & \multicolumn{2}{c|}{all data}                                   & \multicolumn{2}{c|}{coronal hole}  & \multicolumn{2}{c|}{sector reversal}                 & \multicolumn{2}{c}{streamer belt}                     \\
                      & MAPE                          & median  & MAPE                              & median  & MAPE                                 & median  & MAPE                              & median  \\ \hline
$\vp$                & 0.084 & -0.001 & 0.090 & 0.015 & 0.085 & 0.002 & 0.079 & -0.012 \\
$\np$                 & 0.400 & -0.108 & 0.400 & -0.169 & 0.545 & 0.016 & 0.312 & -0.126  \\
$\Tp$                 & 0.327 & -0.078 & 0.291 & 0.016 & 0.382 & -0.347 & 0.316 & -0.033  \\
$\B$                  & 0.277 & -0.053 & 0.222 & -0.023 & 0.352 & -0.107 & 0.265 & -0.052   \\
$\nO$                 & 0.307 & -0.031 & 0.261 & -0.014 & 0.379 & -0.036 & 0.290 & -0.040 \\ \hline
                        & $\chi^2_\text{red}$ & median & $\chi^2_\text{red}$ & median & $\chi^2_\text{red}$ & median & $\chi^2_\text{red}$ & median \\ \hline
$\vp$                 & 50.6 & 22.4 & 59.7 & 24.4 & 47.0 & 24.9 & 47.6 & 19.8   \\
$\np$                 & 37.4 & 3.9 & 80.1 & 3.0 & 19.5 & 6.0 & 22.6 & 3.4    \\
$\Tp$                 & 13.3 & 1.8 & 4.1 & 1.1 & 32.0 & 4.1 & 7.4 & 1.6    \\
$\B$                  & 350.8 & 101.9 & 273.2 & 72.8 & 430.2 & 141.3 & 350.7 & 104.5 \\  
$\nO$                 & 269.0 & 1.0 & 34.0 & 0.3 & 540.8 & 1.8 & 246.2 & 1.6 
\end{tabularx}
\caption{MAPE and $\chi^2_{\text{red}}$ scores for the five reconstructed solar wind parameters $\vp$, $\np$, $\Tp$, $\B,$ and $\nO$. Additionally, the median value for each score is provided. The first column shows the scores for the complete test data set. After applying the scheme from \cite{xu2014new}, the scores of the resulting subsets are recorded in columns two to four.}
\label{tab:MAPE XU}
\end{table*}

\subsection{Dependence on solar wind type }
\label{sec:solar wind type dependence}
Next, we investigate how the reconstruction errors relate to the solar wind type.
Separating the data into the solar wind types as described in \cite{xu2014new} provides clues as to which solar wind type is most difficult to reconstruct. The MAPE scores and the $\chi^2_\text{red}$ scores for each parameter and solar wind type are recorded in Table \ref{tab:MAPE XU}. Additionally, the median value for the underlying distribution of each score is provided. Except for the solar wind proton speed $\vp$, the MAPE scores of the sector reversal solar wind type are consistently worse than those of the coronal hole and streamer belt type. Since the three solar wind types considered here are also affected by different transport effects, the comparison of the reconstruction error in different solar wind types also provides hints as to the influence on transport effects on the reconstruction. Most $\chi^2_\text{red}$ scores follow the pattern that they are higher for higher MAPE scores. Therefore, a poor reconstruction results in comparatively high MAPE scores and $\chi^2_\text{red}$ scores. Nevertheless, there are some exceptions. The proton density $\np$ for sector reversal plasma shows a low $\chi^2_\text{red}$ score compared to coronal hole or streamer belt plasma. We suspect that reconstructing the proton density in coronal hole wind is an indirect result of wave activity. Waves, which at 1AU are mainly observed in coronal hole wind, increase the variability in the proton speed, proton temperature, and the magnetic field strength, while not affecting the proton density. This creates an ambiguity for the proton density, because for the same constant proton density, variable combinations of proton speed, proton temperature, and magnetic field are observed.

Figure~\ref{fig:colage} shows the reconstruction errors for each solar wind parameter ordered by the proton--proton collisional age $\ac$. The proton--proton collisional age (also referred to as the Coulomb number; see \citet{kasper2012evolution}) is well suited to ordering the solar wind observations directly by their collisional history \citep{kasper2019alfvenic,kasper2012evolution,tracy2016constraining}; it also serves as a proxy for the \citet{xu2014new} solar wind types, as discussed in \citep{heidrich2020proton}. Low proton--proton collisional age ($\log\ac\lesssim-0.9$) approximately corresponds to coronal hole wind, intermediate proton--proton collisional age ($-0.9\lesssim\log\ac\lesssim0.1$) to streamer belt plasma, and high proton--proton collisional age ($0.1\lesssim\log\ac$) to sector reversal plasma. Therefore, the three solar wind plasma types can be approximately separated by the proton--proton collisional age.

The locations of the column-wise maxima in Fig.~\ref{fig:colage} and the shape of the distributions in each panel are different for each reconstructed solar wind parameter.
For the magnetic field strength, the core of the coronal hole wind population (low proton--proton collisional age) and the streamer belt population are shifted to the negative relative errors, that is, to overestimation. For the proton density, the peak positions exhibit a similar systematic behaviour to the magnetic field strength. For the proton temperature, the core of the streamer belt population is shifted slightly towards the positive direction, that is, underestimation in the intermediate proton--proton collisional age range, and more strongly in the positive direction in the low proton--proton collisional age range, which again indicates underestimation.  Under the assumption that waves are predominantly observed in coronal hole wind, which is here identified by a low proton--proton collisional age, this systematic change of the positions of the core positions could be interpreted as the effect of the presence or absence of waves on the reconstruction.  The (apparent or real, see \citet{verscharen2011apparent}) heating of the proton core distribution by waves increases the observed proton temperature in coronal hole wind. This is likely not covered well by the neural network model, because in the corresponding proton--proton collisional age range, the distributions of the reconstruction errors for the proton temperature moves to the right, which indicates underestimation of the observed proton temperatures by the neural network model. 

In addition, for the sector reversal plasma, the higher the proton--proton collisional age, the less accurate the reconstruction  for all reconstructed parameters except for the proton speed. This effect is most pronounced in the proton density and in the proton temperature. With compressed slow solar wind from SIRs, the sector reversal plasma contains strongly transport-affected solar wind plasma. In particular, the compression regions affect the proton density, proton temperature, and magnetic field strength, but the proton speed is not affected to the same extent, and the oxygen charge-state ratio is not affected at all. The systematic shifts away from zero, which are visible in the reconstruction error for $\np$, $\Tp$, and $\B$ for the corresponding high proton--proton collisional age in Fig.~\ref{fig:colage}, show that this case is not well represented in the static neural network model. We suspect that the same transport effect is indirectly responsible for the shift of the reconstruction error of the oxygen charge-state ratio in the same high proton--proton collisional age region. Here, the input solar wind parameters, in particular $\np$ and $\Tp$, are systematically changed by the SIR, which makes it more difficult for the static model to recover the transport unaffected charge-state ratio from these strongly transport-affected solar wind parameters. This effect is probably reinforced by the comparatively poorer statistics of sector reversal plasma compared to the other two solar wind types. 

The distribution of the proton speed reconstruction differences is visibly narrower and the MAPE score for the reconstruction of the proton speed is low over  almost the complete range of proton--proton collisional age bins (see bottom-most panel in Fig.~\ref{fig:colage}), which again supports the conclusion that the proton speed reconstruction is the most accurate based on the MAPE score ---although the accuracy is low compared to the small measurement uncertainty of the proton speed; cf. Sect.~\ref{sec:reconstruct}. 

\subsection{Limits of the time-independent model}\label{sec:solarcycle}
The underlying ACE data cover 10 years. This is almost one solar activity cycle. It is well known that the properties of the solar wind all change systematically over time \citep{mccomas2000solar,kasper2012evolution,shearer2014solar}. As our neural network model is stationary, this time-dependent effect cannot be captured by our model. Therefore, in this section, we investigate how strong this effect is and how the reconstruction accuracy varies over time. 

As the Sun and the solar wind properties are less variable during the solar activity minimum, we expected the reconstruction accuracy to worsen with increasing solar activity. Therefore, the best reconstruction would be expected during the solar activity minimum and the worst during the solar activity maximum.

Figure~\ref{fig:test_scores} shows the MAPE score for each reconstructed solar wind parameter and different \citet{xu2014new} solar wind types for each individual time period from the test data set. As discussed in Sect.~\ref{sec:method}, the training, validation, and test data sets are all similarly distributed over time. In each of the four top panels of Fig.~\ref{fig:test_scores}, the MAPE score for each reconstructed parameter is shown for easy comparison. In each panel, we again see that the reconstruction of the proton speed achieves the smallest MAPE scores. The proton density reconstruction shows the largest variation over time, in particular in sector reversal plasma. Linear fits using the mean deviation of the upper and lower bound of the Monte Carlo confidence intervals from each data point  as an estimate for the standard deviation of each value (restricted to 2002-2008) illustrate that, independently of solar wind type, the reconstruction accuracy is similar for all solar wind parameters during all times of the solar cycle. In all cases, the slope of the lines is small, but ---as indicated by the plus and minus symbols to the right of each subplot--- the slopes are significantly different from zero (based on a Wald test for statistical significance). 

A possible explanation for the rising slope ($+$ symbol) of $\B$ and $\nO$ could lie in their respective measurement errors. A rising slope means the reconstruction accuracy is lower during solar activity minimum. The measurement error of $\B$ is given as an absolute error (0.1 nT). Its impact on the MAPE score is greatest for smaller values of $\B,$ which are more likely during solar activity minimum. Similarly, for $\nO$,  less ions are measured during solar activity minimum, and therefore the counting statistics show smaller values, which results in greater measurement uncertainties.
As discussed in Sect.~\ref{sec:solar wind type dependence} and shown in Table \ref{tab:MAPE XU}, the sector reversal plasma reconstruction is the least accurate. In Fig.~\ref{fig:test_scores} sector reversal plasma shows the greatest variability between the different data points (each data point corresponds to approximately one Carrington rotation). The outliers in sector reversal plasma are also the most extreme. One explanation could be that the \citet{xu2014new} scheme employed here misidentifies some coronal hole plasma as sector reversal plasma (and vice versa). As plasma from stream interaction regions is strongly affected by transport effects, this regime is  more difficult to reconstruct and therefore has a negative affect on the reconstruction accuracy of whichever group it gets sorted into; in this case, sector reversal plasma.

However, in contrast to our expectations, for all reconstructed parameters, the influence of the solar activity cycle is small compared to the uncertainty arising from the measurement uncertainty. Therefore, with the large underlying measurement uncertainties on the proton temperature, the proton density, and the oxygen charge-state ratio, the reconstruction is not accurate enough to allow a considerably better reconstruction during the solar activity minimum. As the measurement uncertainty on the proton temperature, the proton density, and the oxygen charge-state ratio is also large compared to the systematic variations of these parameters with the solar activity cycle, this is not surprising.

\section{Discussion}\label{sec:conclusion}

The properties of the solar wind are determined by a combination of the conditions in the solar source region and the transport history experienced by the solar wind. As a result, the proton plasma properties, the magnetic field strength, and the charge-state composition are correlated to each other \citep{lepri2013solar,mccomas2000solar,vonSteiger2000}. Here, we investigate whether a combination of four of these solar wind parameters is sufficient to reconstruct the remaining fifth solar wind parameter. If the considered solar wind parameters were to contain all the information necessary, a perfect reconstruction would be possible. We therefore consider the obtained reconstruction accuracy as a surrogate to quantify the degree to which other (unknown) hidden parameters play a role in the correlations between our considered solar wind parameters. By analysing how this changes under different solar wind conditions with different respective transport histories, we can extend this argument to the impact of the varying influence of transport processes on the properties of the solar wind.

To this end, we investigate interdependencies between different solar wind parameters, namely $\vp$, $\np$, $\Tp$, $\B$, and $\nO$. All five considered solar wind parameters depend on the respective solar source of the observed solar wind. However, they are affected differently (or not at all, as in the case of $\nO$) by different transport effects, which obscures the original source-driven interrelationships. We use a neural network as a general function approximator to model the interdependencies of these solar wind parameters.

The lowest mean absolute reconstruction error is achieved for the proton speed $\vp(\Tp, \np, \B, \nO)$. One possible interpretation is that the information carried by the proton speed can be extracted from the other four parameters. However, the proton speed is also associated with a small measurement uncertainty, and compared to the measurement uncertainty the accuracy of the proton speed reconstruction is lower than for the proton density and proton temperature.
Nevertheless, our results indicate that the solar wind parameter proton speed can be replaced by other measurements, and appears therefore the least important parameter to measure. Reconstruction of any of the other four parameters, $\Tp$, $\np$, $\B$, or $\nO$, has proven to be more difficult. Here, the (absolute) reconstruction errors remain high, $\approx 30 \%$. The accuracy of the proton density, proton temperature, and the oxygen charge-state ratio is strongly limited by the underlying measurement uncertainty. On the one hand, the measurement uncertainty determines the accuracy of the reconstructed parameter itself. This is the case for, for example, the proton temperature, which shows a high (and therefore poor) MAPE score, but reaches the lowest (best) $\chi^2_{\text{red}}$  score. On the other hand, the large measurement uncertainties of, for example, the proton temperature and the oxygen charge-state ratio also limit the reconstruction of all other parameters in our setup, because inaccurate input parameters also affect the output parameter. We suspect this effect is the reason for the low reconstruction accuracy compared to the measurement uncertainty we obtained for the proton speed and the magnetic field strength.   For the oxygen charge-state ratio, the average reconstruction accuracy compared to the measurement uncertainty is very low, but  the reconstruction accuracy is on the order of the measurement uncertainty for
the majority of the individual data points. Therefore, measuring these quantities with higher accuracy is important in order to understand the interdependencies of the solar wind parameters. This is of particular interest in the case of the  oxygen charge-state ratio $\nO$. As the only parameter not affected by transport effects, this latter contains unique information that cannot be substituted by a non-linear relationship between the proton plasma parameters with high absolute accuracy, and the reconstruction of all solar considered solar wind parameters is likely inhibited by the large measurement uncertainties on $\nO$. This emphasises the need for heavy ion instruments, such as ACE/SWICS, or the Heavy Ion Sensor (HIS), which is part of the Solar Wind Analyzer (SWA) \citep{owen2020solar} on board Solar Orbiter. These instruments require a stable high-voltage supply, which is challenging to design, and analysis of the data they provide is a complex undertaking. However, our results illustrate that the effort to build instruments of this type is necessary. %

From the point of view of the solar source region, our reconstruction approach implies that the proton speed carries less detailed information about the source conditions than the other four solar wind parameters. Studying the signatures in $\Tp$, $\np$, $\B$, and $\nO$ may therefore provide a better chance to capture relevant details of a specific solar source region and the solar wind release mechanisms. Given the large measurement uncertainties on $\Tp$, $\np$, and $\nO$, these four solar wind parameters cannot be completely reconstructed from each other, and therefore investigating which other properties ---not included in this study--- determine their variability may help us to identify the underlying mechanisms behind the release of (slow) solar wind.

The reconstruction accuracy differs depending on solar wind type. Based on  the absolute reconstruction accuracy (estimated with the MAPE score) and the reconstruction accuracy relative to the measurement uncertainty (estimated with the $\chi^2_{\text{red}}$ and the median of the individual $\chi^2$ scores), the reconstructions of $\B$, $\Tp$, and $\nO$ are best in coronal hole wind, the reconstruction of $\vp$ is best in sector reversal plasma, and the reconstruction of $\np$ is best in streamer belt plasma. This illustrates that reconstructions face different challenges in different solar wind types, which differ both in the properties of the solar source region and in the transport history experienced during the solar wind travel time.

To further investigate our results from the point of view of the transport history of solar wind, we make use of the proton--proton collisional age, which ---although not defined for this--- can serve as a proxy to differentiate between solar wind types with different transport histories \citep{heidrich2020proton}. Coronal hole wind is often influenced by wave activity. Waves have several effects on the solar wind plasma: The core of the proton population is (or is apparently) heated, probably preferentially perpendicular to the magnetic field; waves are speculated to play a role in the formation of the beam \citep{marsch1982wave,verniero2020parker,d2015origin,panasenco2020exploring,louarn2021multiscale}; and wave--particle interaction likely plays a role in differential streaming \citep{kasper2012evolution,janitzek2016high,marsch1982wave}. Here, we cannot resolve these different effects, but argue that they are all mainly confined to coronal hole wind (or Alfv{\'e}nic slow solar wind), which is
typically associated with low proton density, high proton temperature, and high solar wind speed, which all lead to a low proton--proton collisional age. Therefore, we assume that the effect of waves as transport processes are relevant in solar wind with a low proton--proton collisional age. In this regime, we find that our neural network reconstruction tends to underestimate the proton temperature and (to a lesser degree) the magnetic field strength. Among the solar wind parameters considered here, these two, $\Tp$ and $\B,$ are exactly the parameters that are expected to be most influenced by Alfv{\'e}n waves. Therefore, our neural network reconstruction appears to focus on the underlying (source-driven) relationship between the solar wind parameters and not on the effect of waves on the solar wind plasma. Indirectly, this is supported by the observation that the oxygen charge-state ratio does not show a preferential over- or underestimation in the coronal hole wind regime. This is in agreement with the expectation that the oxygen charge-state ratio is not affected by transport effects.

If we focus on solar wind with a particularly high proton--proton collisional age, this selects solar wind with high proton densities, high proton temperatures, and low proton speeds. These conditions are best realised in compressed slow solar wind in SIRs and in the preceding slow solar wind. Therefore, as argued in \citet{heidrich2020proton}, we consider solar wind with a high proton--proton collisional age as a proxy for SIRs, which tend to be included in the sector reversal plasma category in the \citet{xu2014new} categorisation. The sector reversal plasma also included a very slow, cool, and dense solar wind type \citep{sanchez2016very}, which also falls in the high proton--proton collisional age regime. In this regime, we observe a systematic increase in the overestimation of the proton temperature and a systematic increase in the underestimation of the proton density. The underestimation of the proton densities again implies that the neural network model is probably tailored to `normal' conditions that are unaffected by transport, and is therefore ill-equipped to adapt to the different relationship between proton density and proton temperature in compressed solar wind. For the proton temperature, the observed underestimation is the result of a compromise between attempting to model the higher proton temperatures in compression regions and those in the very slow solar wind identified in \citet{sanchez2016very}. This bolsters the argument that the information contained in the solar wind parameters considered here (and in other studies) is likely not sufficient to completely characterise the plasma, its solar source, or the experienced transport history. Our results also support the idea that any solar wind classification using only one or a few of the solar wind parameters considered here contains an inherent bias towards greater accuracy during certain conditions that are more or less affected by transport processes.

Our neural network models a static, time-independent relationship between the considered solar wind parameters. However, all considered solar wind parameters systematically change with the phase of the solar activity cycle.  Therefore, in principle, our neural network model cannot be expected to perform equally well in all phases of the solar activity cycle. However, investigating how the reconstruction accuracy changes over (almost) one solar cycle shows that the influence of the high measurement uncertainties on the underlying parameters is stronger than a potential solar activity cycle-dependent effect.  That the model achieves better results for the reconstruction of the oxygen charge-state ratio during the solar activity maximum is probably also an effect of the measurement uncertainty on $\nO$. During the solar activity minimum phase, observations of very dilute plasma are more likely. This condition can lead to very low count rates in ACE/SWICS and therefore to a very high measurement uncertainty.

Although the oxygen charge-state ratio is the only solar wind parameter considered here that is not affected by any transport effects that complicate the relationship between the different considered solar wind parameters, the neural network reconstruction of the oxygen charge-state ratio does not prove to be easier than that of the other transport-affected solar wind parameters. This can be caused by at least three mechanisms: (1) The oxygen charge-state ratio at the source depends (also) on a property that is not included in our analysis; (2) recovering sufficiently detailed information on the solar source region (which determines the oxygen charge-state ratio) from the proton plasma properties and the magnetic field strength is hindered by the influence of the transport history, which strongly affects the proton temperature and the proton density; and (3) the high measurement uncertainties are unconducive to a good reconstruction.

\section{Conclusion}

We investigated non-linear relationships between different solar wind parameters; namely proton speed, proton density, proton temperature, magnetic field strength, and the oxygen charge-state ratio. 
Our findings suggest that only the proton speed can be substituted with other measurements with reasonable absolute and relative accuracy. This implies that the proton speed carries less unique information about the solar source region and transport effects than the other considered solar wind parameters. The precision of the reconstructions of the proton density, proton temperature, and the oxygen charge-state ratio is constrained by their respective measurement uncertainties. While the average reconstruction accuracy of the oxygen charge-state ratio compared to the measurement uncertainty is generally low, most individual data points exhibit reconstruction accuracy in line with the measurement uncertainty. While the magnetic field strength can be measured with high accuracy, its reconstruction in our study is similarly inhibited by the comparatively high uncertainties on proton density, proton temperature, and the oxygen charge-state ratio. Therefore, to further our understanding of the relationships between different solar wind parameters and the process they originate from, it is crucial to further enhance the measurement accuracy for these quantities.
  
Our neural network reconstruction appears to focus on the underlying relationship driven by the sources of the solar wind, rather than disentangling the impact of transport effects such of wave-particle interactions, collisions, or compression regions on the solar wind plasma. Nevertheless, the reconstruction accuracy clearly differs depending on the solar wind type. We note that different transport effects are dominant in different respective solar wind types, and therefore transport effects, such as wave--particle interactions in cornal hole wind, collisions in slow solar wind, and compression regions in SIRs, limit the potential accuracy of identifying the source region of the solar wind purely based on the observations of proton speed, proton density, proton temperature, and magnetic field strength. Our results therefore underline the importance of measuring the charge states of the solar wind directly and with high accuracy.
  
For complex models based on magnetohydrodynamics and solar corona magnetic field models \citep{arge2000improvement,cranmer2005,cranmer2007self,van2010data, pizzo2011wang, schultz2011space,van2014alfven,pomoell2018euhforia}, capturing the properties of SIRs tends to be comparatively difficult. The fact that our simple approach of ad hoc neural network reconstruction is also least accurate for the solar wind type that contains the most SIRs suggests that the additional effect of compression ---which dominates the plasma properties in SIRs--- needs to be considered in the design of highly accurate models. Consequently, incorporating comprehensive details regarding transport effects, compression regions, and the progressive impingement of faster solar wind into SIRs \citep{hofmeister2022area} into consistent MHD-driven solar wind models holds promise for enhancing their accuracy.

\begin{acknowledgements}
  This work was supported by the \emph{Deutsches Zentrum für Luft-
      und Raumfahrt} (DLR) as SOHO/CELIAS 50 OC 2104.  We further
      thank the science teams of  ACE/SWEPAM, ACE/MAG as well as
      ACE/SWICS for providing the respective level 2 and level 1 data
      products.
      The Sunspot data is taken from the World Data Center SILSO, Royal Observatory of Belgium, Brussels.
\end{acknowledgements}
%
\bibliographystyle{aa} 
\bibliography{aa} 
%

\end{document}